\documentclass{article}

\usepackage[preprint]{neurips_2022}

\usepackage[utf8]{inputenc} % allow utf-8 input
\usepackage[T1]{fontenc}    % use 8-bit T1 fonts
\usepackage{hyperref}       % hyperlinks
\usepackage{url}            % simple URL typesetting
\usepackage{booktabs}       % professional-quality tables
\usepackage{amsfonts}       % blackboard math symbols
\usepackage{nicefrac}       % compact symbols for 1/2, etc.
\usepackage{microtype}      % microtypography
\usepackage{xcolor}         % colors

\usepackage{amsmath,amssymb,amsfonts}
\usepackage{multirow}
\usepackage{psfrag}
\usepackage{subfigure}
\usepackage{wrapfig}
\usepackage{color}
\usepackage{soul}
\usepackage{xcolor}
\usepackage{lipsum}
\usepackage[ruled,vlined,linesnumbered]{algorithm2e}
\usepackage{mathrsfs}
\usepackage{pbox}
\usepackage{amsfonts}
\usepackage{mathtools}
\usepackage{amsthm}
\usepackage{microtype}

\graphicspath{{Figures/}}

\title{LSTM-Based Adaptive Vehicle Position Control for Dynamic Wireless Charging}

\author{%
  Lokesh Chandra Das\\
  Department of Computer Science\\
  University of Memphis\\
  Memphis, TN \\
  \texttt{ldas@memphis.edu} \\
  \And
   Dipankar Dasgupta\\
  Department of Computer Science\\
  University of Memphis\\
  Memphis, TN \\
  \texttt{ddasgupt@memphis.edu} \\
  \And
  Myounggyu Won\\
  Department of Computer Science\\
  University of Memphis\\
  Memphis, TN \\
  \texttt{mwon@memphis.edu} \\
}

\begin{document}

\maketitle

\begin{abstract}

Dynamic wireless charging (DWC) is an emerging technology that allows electric vehicles (EVs) to be wirelessly charged while in motion. It is gaining significant momentum as it can potentially address the range limitation issue for EVs. However, due to significant power loss caused by wireless power transfer, improving charging efficiency remains as a major challenge for DWC systems. This paper presents the first LSTM-based vehicle motion control system for DWC designed to maximize charging efficiency. The dynamics of the electromagnetic field generated by the transmitter coils of a DWC system are modeled based on a multi-layer LSTM. The LSTM model is used to make a prediction of the lateral position where the electromagnetic strength is expected to be maximal and to control the EV motion accordingly to optimize charging efficiency. Simulations were conducted to demonstrate that our LSTM-based approach achieves by up to 162.3\% higher charging efficiency compared with state-of-the-art vehicle motion control systems focused on keeping an EV in the center of lane. 

\end{abstract}

\section{Introduction}
\label{sec:introduction}

Energy security is one of the highest priorities for many countries. The largest energy consumer is the transportation industry. For instance, in U.S., the transportation industry has consumed more than 28\% of the total energy in just one year~\cite{patil2017wireless}. Especially, gasoline fuel consumption mostly for vehicles accounts for more than 56\% from the total transportation energy consumption~\cite{patil2017wireless}. Electric vehicles (EVs) have received significant attention as a viable solution to reduce energy consumption and alleviate environmental effects. However, developing inexpensive, durable, and reliable energy storage system remains as a significant challenge. In fact, one of the primary concerns that discourage purchase of an EV for many consumers is the range anxiety and long charging time~\cite{guidi2020dynamic}. 

Dynamic wireless charging (DWC) is an emerging technology developed to address the range anxiety and long charging-time problems~\cite{patil2019coil, ahmad2017comprehensive}. The DWC technology allows vehicles to be charged wirelessly while in motion based on the magnetic coupling between the coils embedded under the road surface and those fit in an EV~\cite{lukic2013cutting}. The transmitter coils positioned under the road are provided with a high-frequency current and generate electromagnetic field, which is picked up by the EV to charge the EV battery~\cite{ahmad2017comprehensive}. The first concept of DWC was designed by Bolger \emph{et al.}~\cite{bolger1978inductive} in 1978. Extensive research has been conducted to develop numerous DWC solutions~\cite{guidi2020dynamic,mi2016modern,ahmad2017comprehensive}. 

To make DWC widely accepted as a convenient, reliable, and flexible solution, there are significant challenges to be addressed~\cite{guidi2020dynamic}. One of them is the misalignment between the transmitter and receiver coils as the power transfer efficiency depends on accurate coil alignment~\cite{moon2014analysis,van2011control,panchal2018review,chen2015promoted,hwang2017autonomous}. Since EVs are charged while in motion, such misalignment is unavoidable, causing significant fluctuation of power transfer efficiency~\cite{tian2020vision}. 

Numerous works have been proposed to mitigate the misalignment problem. Hardware-based solutions are focused on modifying~\cite{chen2016cost,chow2014investigation} or adding~\cite{shin2013design,kim2013coil} a new hardware component, \emph{e.g.,} changing the geometry or configuration of the coil. Tracking-based approaches are, on the other hand, designed to align the EV in the center of lane based on a vehicle positioning algorithm~\cite{byun2015vehicle,hwang2017autonomous}. Vision-based solutions that use a camera to track the lateral position of EV and position it in the center of lane have been studied~\cite{tian2020vision}. However, a critical limitation of these solutions is that the dynamics of electromagnetic field generated by the transmitter coils of the DWC system are not taken into account to control the vehicle motion, potentially leading to degraded power transfer efficiency. 

In this paper, we present a LSTM-based adaptive vehicle motion control system for DWC. To optimize the power transfer efficiency, the lateral position of an EV is adaptively adjusted in response to the dynamics of the electromagnetic field generated by the transmitter coils of a DWC system. We design a multi-layer LSTM network that effectively captures the dynamics of the electromagnetic field. The LSTM model allows the EV to predict the optimal lateral position where the power transfer efficiency is expected to be maximized and control its motion accordingly. Through simulations, we, for the first time, uncover the strong potential of the machine learning-assisted approach for DWC to significantly boost the power transfer efficiency. More specifically, compared with a state-of-the-art approach focused on keeping the EV in the center of lane to improve the power transfer efficiency, we demonstrate that our LSTM-based approach leveraging prediction of the optimal lateral position where the electromagnetic strength is maximized enhances the power transfer efficiency by up to 162.3\%. 

This paper is organized as follows. In Section~\ref{sec:related_work}, we thoroughly review solutions designed to address the misalignment problem for DWC. We then perform a motivational study to better understand the dynamics of the electromagnetic field generated by the transmitter coils in Section~\ref{sec:motivation}. And then, we present the details of the design of the proposed LSTM model in Section~\ref{sec:main}. The simulation results are presented in Section~\ref{sec:results} followed by conclusion and future work in Section~\ref{sec:conclusion}.

\section{Related Work}
\label{sec:related_work}

This section reviews the literature dedicated to addressing the misalignment problem for DWC. We categorize those solutions largely into hardware-based, tracking-based, and vision-based approaches.

Hardware-based approaches are characterized by modifying the hardware of the DWC system. Chen \emph{et al.} propose a novel geometry of coils to improve power transfer efficiency~\cite{chen2016cost}. Chow \emph{et al.} consider placing multiple coils in an orthogonal configuration~\cite{chow2014investigation}. Kalwar \emph{et al.} combine multiple coils of different geometry into a single unit~\cite{kalwar2016coil}. Some hardware-based approaches attempt to add a new hardware component. For instance, E-shape or U-shape ferrite cores are integrated with the source coils~\cite{shin2013design,kim2013coil}. Active coil resonance frequency tuning circuits are applied~\cite{gao2015uniform,hu2016frequency}. A novel arrangement method for sensing coils is developed to detect lateral misalignment problem~\cite{tavakoli2021cost}. Although these hardware-based approaches improve the power transfer efficiency, the dynamics of the electromagnetic field are largely ignored. Also, some approaches have limitations in terms of the space and weight constraints for specific environments~\cite{hwang2017autonomous}.

Tracking-based approaches are focused on monitoring the alignment between EV and DWC system using a vehicle position tracking system. The global positioning systems (GPS) has been widely adopted for tracking the EV position. However, due to the high positioning error of GPS, most tracking-based approaches utilize different types of sensors for tracking the EV position. For instance, a radio-frequency identification (RFID) tag or a magnetic marker has been adopted~\cite{ryu2015wireless,shuwei2014research,choi2009autonomous}. However, since the strength of the magnetic field decays rapidly as the distance between the detector and the marker increases, those magnetic markers or RFID tags must be placed very close to each other to achieve high-resolution tracking performance, thereby increasing the construction cost significantly. In an effort to address this problem, sensor systems with greater range have been integrated with the charging system~\cite{xu2009magnetic}. Additionally, a Gaussian function-based algorithm is developed to increase the detection accuracy~\cite{byun2015vehicle}. However, these tracking system-based approaches do not work effectively when the vehicle speed is very high since the magnetic field changes dynamically as the vehicle travels fast~\cite{hwang2017autonomous}.

The inherent limitations of tracking-based approaches have been addressed by developing the autonomous coil alignment system (ACAS). The main advantage is that it is designed to track the vehicle's misalignment position based only on the voltage changes in the vehicle's load coil~\cite{hwang2015autonomous}. An enhanced version of ACAS has been introduced to reduce the complexity of the algorithm and hardware~\cite{hwang2017autonomous}, allowing for wider applicability with DWC systems with varying specifications. ACAS not only addresses the challenges of external sensor-based approaches since it is implemented on top of the existing DWC system, but also reduces the implementation cost. Although ACAS is demonstrated as an effective solution for addressing the misalignment problem, it is only compatible with specific DWC systems. Furthermore, ACAS does not account for the effect of dynamically changing electromagnetic field of the transmitter coils on the power transfer efficiency~\cite{hwang2017autonomous}.  

%Sensor-less approaches have also been studied. Sonapreetha \emph{et al.} propose a sensor-less approach that detects the EV position based on the phase angle ``by which current lags the voltage reflected from the edge position of vehicle''~\cite{wang2017analysis, patil2019coil} \hl{check}. Tang \emph{et al.} propose to achieve high power transfer efficiency by ``tuning Vin to make dD/dVin (Vin is the DC input for the primary inverter, and D is the duty cycle of the secondary converter) equal to a constant coefficient determined by the system parameters.''~\cite{tang2017low} \hl{check}. 

A vision-based approach has been proposed recently that uses a camera to track the lateral position of EV to keep the vehicle in the center of lane to improve power transfer efficiency~\cite{tian2020vision}. However, due to the dynamically changing electromagnetic field, positioning the EV in the center of lane does not guarantee optimal power transfer efficiency. In contrast to existing works, our solution is the first machine learning-assisted approach that adaptively controls the EV position based on the prediction of the optimal position with maximum electromagnetic strength by effectively modeling the dynamics of the electromagnetic field.  

%Another distinctive aspect of our solution is the prediction of the electomagnetic field to control the laterla position of the vehicle before it reaches at the target point so the vehicle is fully aligned at the optimal position. This way, the power transfer efficiency can be significantly improved. 

\section{Dynamics of Electromagnetic Field in DWC}
\label{sec:motivation}

We conduct simulations to demonstrate the dynamically changing electromagnetic field generated by the transmitter coils in space and time domains and its effect on power transfer efficiency. We use QuickField~\cite{quickfield}, an electromagnetic field simulation platform designed for various types of applications for electrical, thermal, bio-, and chemical engineering. This simulation platform is suitable for this motivational study as it allows us to simulate the time-varying electromagnetic field generated by non-linear coils under varying conditions.

\begin{wrapfigure}{r}{0.46\columnwidth}
	\vspace{-20pt}
	\begin{center}
		\includegraphics[width=0.6\linewidth]{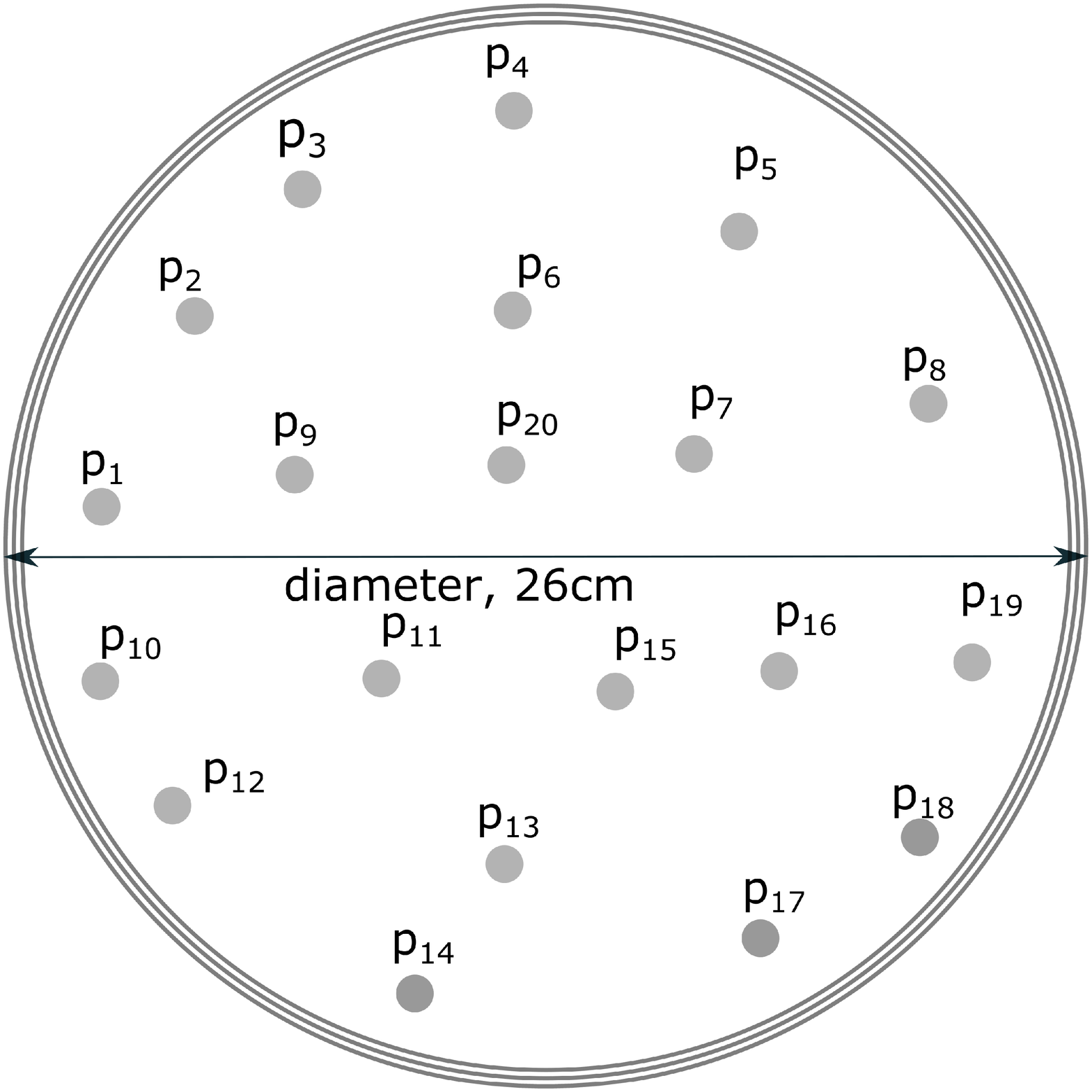}
		\caption{The 20 points on the transmitter coil where electromagnetic strength is measured.}
		\label{fig:tx_coils}
	\end{center}
\end{wrapfigure}

A general setting for a DWC system is considered, where the transmitter coils are installed under the ground, and the receiver coils are mobile as they are mounted on the EV chassis. More specifically, the transmitter coils are at 3cm depth under the ground, and the shape of the coils (\emph{i.e.}, transmitter and receiver coils) is circular with a radius of 13 cm. As shown in Fig.~\ref{fig:tx_coils}, a total of 20 points are randomly and uniformly selected. We measure electromagnetic strength at each point when the receiver coils pass over the transmitter coils. The unit of measurement for electromagnetic strength is $A/m$ (\emph{i.e.,} Ampere per meter). More precisely, the unit for magnetic permeability $\mu$ is defined as $N/A^2$ where $N$ is Newton, and the magnetic field $B$ is measured in Tesla which is $N/Am$. The magnetic field strength is then measured in $B/\mu = T/(N/A^2) = (N/Am)/(N/A^2) = A/m$.

\begin{figure}[h]
	\centering
	\begin{minipage}[b]{0.45\textwidth}
		\centering
		\includegraphics[width=\linewidth]{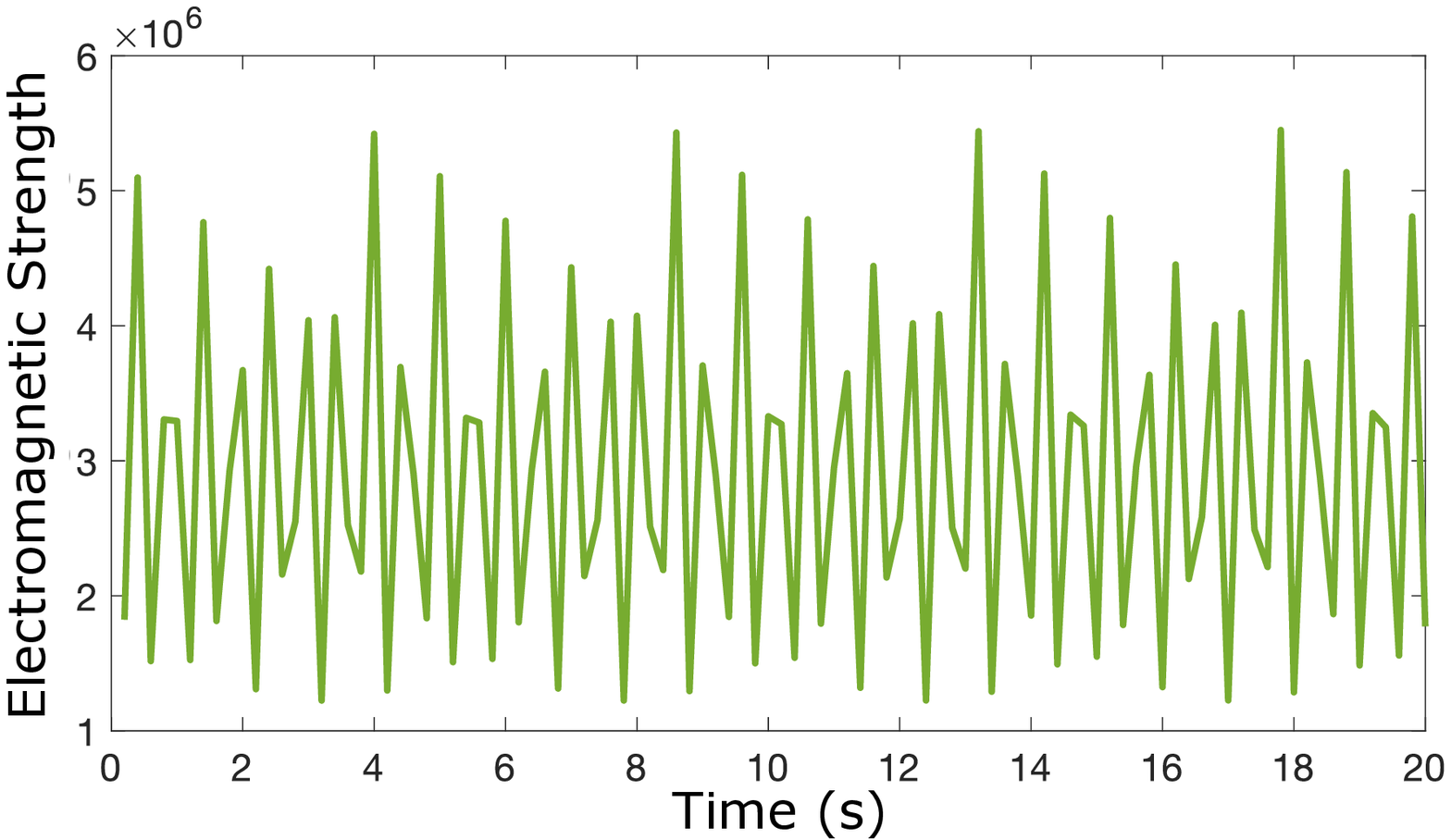}
		\caption {Electromagnetic strength measured at a point over time demonstrating dynamically changing electromagnetic strength.}
		\label{fig:motivation3}
	\end{minipage}
 	\hspace{5mm}
	\begin{minipage}[b]{0.48\textwidth}
		\centering
		\includegraphics[width=\linewidth]{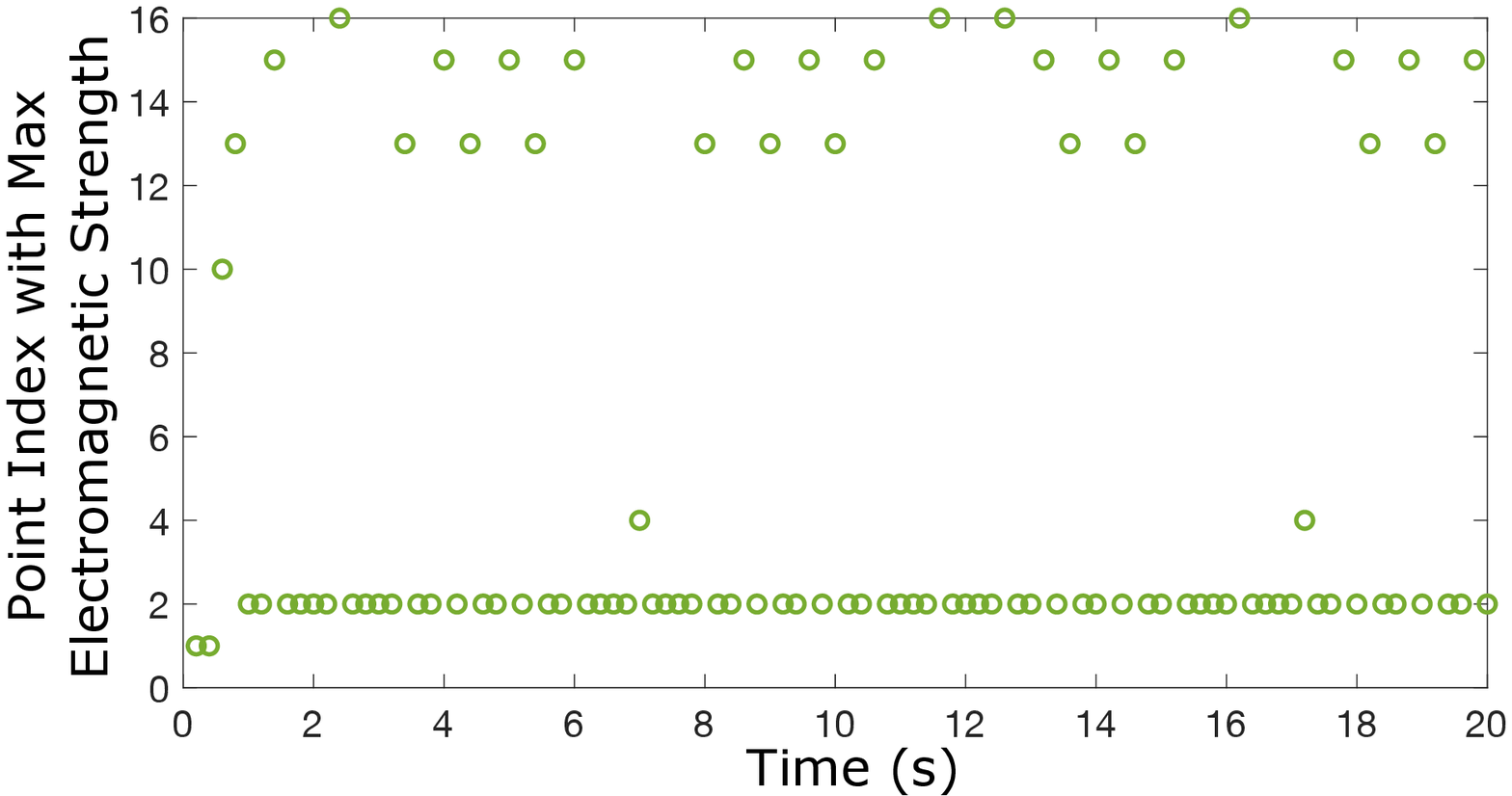}
		\caption {Point indices with maximum electromagnetic strength, which dynamically change over time.}
		\label{fig:motivation4}
	\end{minipage}
\end{figure}

Fig.~\ref{fig:motivation3} depicts electromagnetic strength measured over a period of 20 seconds at a randomly selected point. The figure demonstrates the highly dynamic nature of the electromagnetic field generated by the transmitter coils. The percentage difference between the lowest and highest electromagnetic strength measured at this particular point is over 85\%. Another interesting observation is that the peak electromagnetic strength is not always observed at the center of the field. Fig.~\ref{fig:motivation4} depicts the point index with the maximum electromagnetic strength measured with an interval of 1 second. The results show that a point with maximum electromagnetic strength dynamically changes over time. Consequently, this simulation study motivates us to develop an effective method for predicting a point with maximum electromagnetic strength and align the lateral position of EV with that point before the EV passes over the transmitter coils to maximize the power transfer efficiency.

\section{LSTM-Based Adaptive Vehicle Position Control for DWC}
\label{sec:main}

In this section, we present the details of our LSTM-based adaptive vehicle position control system for DWC. We aim to model the dynamics of the electromagnetic field using LSTM and control the lateral position of the EV to optimize the power transfer efficiency. We start with an overview of the proposed system (Section~\ref{sec:system_overview}). We then describe our dataset (Section~\ref{sec:dataset}) and the details of our LSTM model  (Section~\ref{sec:LSTM}).

\subsection{System Overview}
\label{sec:system_overview}

\begin{figure}[h]
	\centering
	\includegraphics[width=.8\columnwidth]{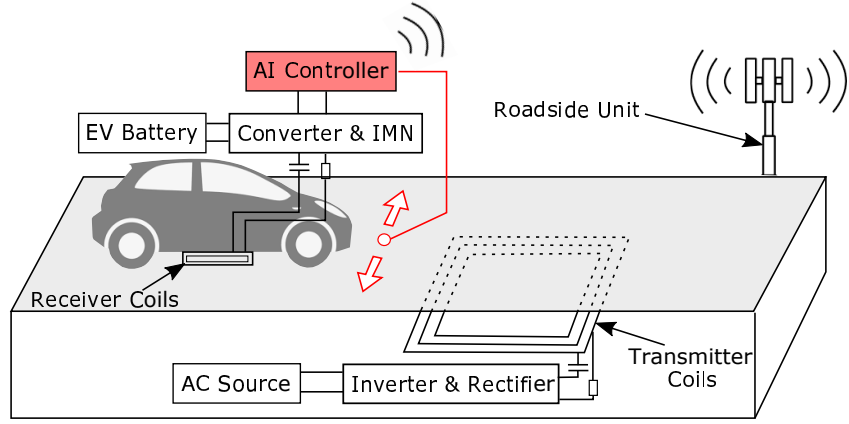}
	\caption {An architecture of the dynamic wireless charging system integrated with the proposed LSTM-based vehicle position control solution.}
	\label{fig:overview}
\end{figure}

An overview of a DWC system integrated with our LSTM-based vehicle motion control solution is presented. There are four main components: power transfer module, power receiver module, EV motion controller, and roadside unit (RSU). The contribution of this paper is focused on designing the EV motion controller. Fig.~\ref{fig:overview} depicts the system architecture of the DWC system. The power transfer module is installed beneath the road surface. The power receiver module is mounted on the EV chassis. The electric power is provided to the EV's battery via magnetic coupling resonance between the transmitter and receiver coils. The RSU is responsible for collecting the electromagnetic field data to train the LSTM model. The EV motion controller sends its speed information to the RSU via vehicle-to-everything (V2X) communication. The RSU then estimates the arrival time of the EV, based on which the RSU computes the EV's optimal lateral position using the LSTM model. The RSU then sends the optimal lateral position to the EV motion controller via V2X. The EV motion controller adjusts the vehicle motion to align the vehicle with the optimal lateral position to maximize the power transfer efficiency.

\subsection{Dataset Preparation}
\label{sec:dataset}

A dataset used to train and test our LSTM model is explained in this section. We use the same simulation setting described in Section~\ref{sec:motivation} to collect the dataset. We measure the electromagnetic strength at the 20 points every 200ms. A point that has the largest electromagnetic strength is recorded and used as a feature vector. More specifically, our feature vector $\overrightarrow{x_i}$ is $[T_i \mbox{ } A_i \mbox{ } p_x \mbox{ } p_y]$ where $T_i$ is the $i$-th time step (the interval of time step is 200ms), $A_i$ is the electromagnetic strength, $p_x$ and $p_y$ are the coordinates of the point with maximum electromagnetic strength. Consequently, our raw dataset consists of a sequence of feature vectors, $\{ \overrightarrow{x_1}, ..., \overrightarrow{x_N}\}$.

\begin{figure}[h]
	\centering
	\includegraphics[width=.7\textwidth]{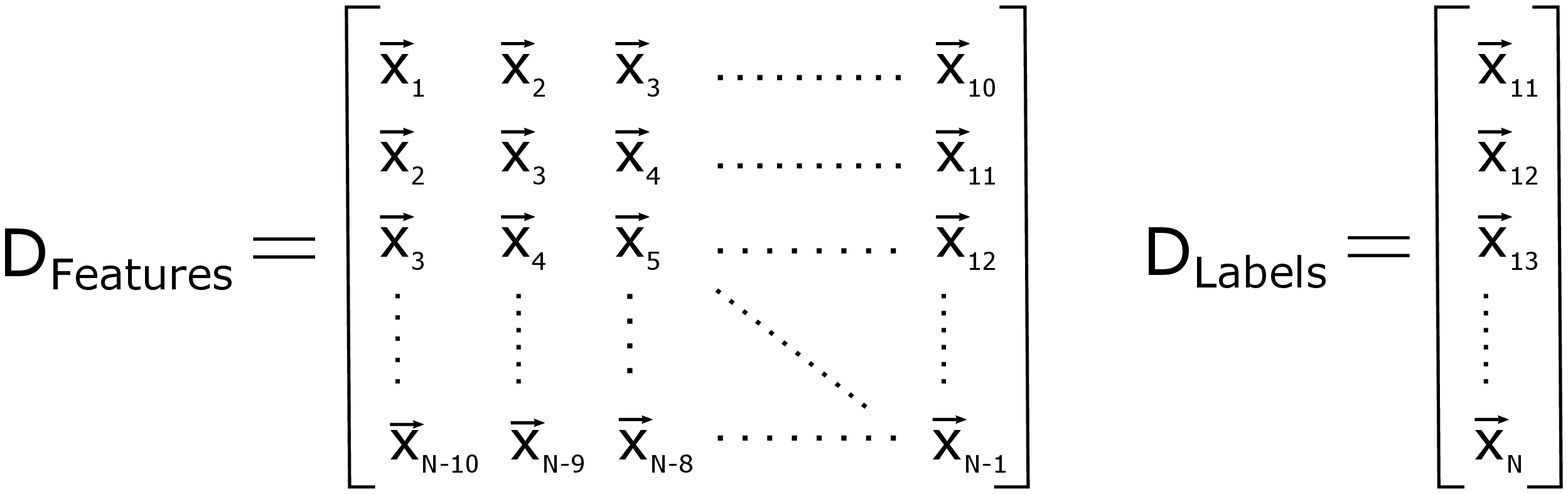}
	\caption {The preprocessed dataset consisting of features and labels prepared based on the sliding window-based method for training and testing our LSTM model.}
	\label{fig:timeseries_dataset}
\end{figure}

The raw dataset is preprocessed to make it suitable for training and testing our LSTM model. We use standard MinMaxScaler to scale up all feature values between 0 and 1 to allow our LSTM model to converge faster. A sliding window method is applied to restructure the raw dataset. More specifically, a sliding window of size 10 comprising of 10 successive feature vectors is used to predict the point with the largest electromagnetic strength in the next time step. The sliding window moves forward by one position. Fig.~\ref{fig:timeseries_dataset} shows an example of converted dataset consisting of a feature matrix $D_{Features}$ and the corresponding label matrix $D_{Label}$. The $i^{th}$ row of $D_{Features}$ represents 10 consecutive feature vectors used to predict the point with the largest electromagnetic strength in the next step which is specified in the $i^{th}$ row of $D_{Label}$.

\subsection{LSTM Model Design}
\label{sec:LSTM}

%The Recurrent Neural network (RNN) is helpful in sequential data, i.e., time series ~\cite{graves2013speech}, because of its feedback loops to iterate over timesteps of a sequence and internal memory to store the information it already experienced. However, vanilla RNN is unable to keep long-term dependencies of the data and thus suffers from vanishing gradient problem.

%\begin{figure}[h]
%\centering
%\includegraphics[width=.99\columnwidth]{RRN_unrolled.eps}
%\caption {The structure of RNN}
%\label{fig:rnn_unrolled}
%\end{figure}%

\begin{figure}[t]
	\centering
	\includegraphics[width=.9\textwidth]{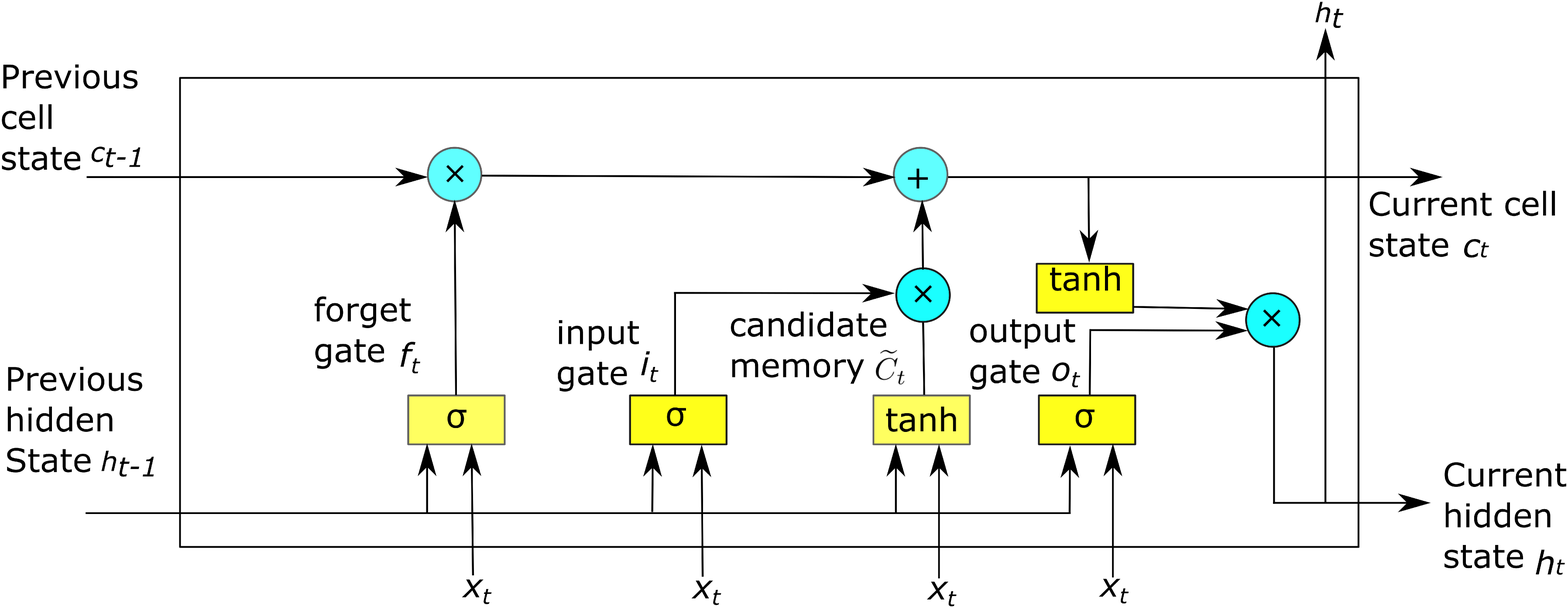}
	\caption {A structure of an LSTM cell.}
	\label{fig:lstmcell}
\end{figure}

Long Short-Term Memory networks (LSTMs) proposed by~\cite{schmidhuber1997long} are designed to learn long-term dependencies and effectively deal with the vanishing gradient problem~\cite{hou2019normalization,li2019lstm}. LSTM is highly suitable for applications that make predictions based on time series data~\cite{cao2020spectral}. The memory block is the basic unit of the LSTM architecture. Each memory block has one or more memory cells and the input, output, and forget gates which all are shared by each cell in the memory block of the LSTM. Fig.~\ref{fig:lstmcell} displays the internal structure and operations of a LSTM cell. The forget gate $f_t$ controls which information is to be kept and which is to be thrown away based on the current input $x_t$ and the previous hidden state $h_{t-1}$, passed through the sigmoid activation function $\sigma$ (Eq.~\eqref{eq:forget}). The input gate $i_t$ decides which information is necessary for the current state and which is not relevant based on the sigmoid activation function (Eq.~\eqref{eq:forget}).

\begin{equation}
	f_t = \sigma(W_h^fh_{t-1} + W_i^fx_t +b_f), \mbox{ } i_t = \sigma(W_h^ih_{t-1} + W_i^ix_t + b_i),
	\label{eq:forget}
\end{equation}

where $W_h^*$ and $W_i^*$ are the weight matrices for the previous hidden state $h_{t-1}$ and current input vector $x_t$ of corresponding gates, and $b_f$ and $b_i$ are biases. An intermediate candidate value matrix $\widetilde{C}_t$ is computed based on the $tanh$ activation function (Eq.~\eqref{eq:candiate}). Using the forget gate $f_t$, input gate $i_t$, and candidate value matrix $\widetilde{C}_t$, the old cell state $c_{t-1}$ is updated with the new cell state $c_t$ (Eq.~\eqref{eq:candiate}). 

\begin{equation}
	\widetilde{C}_t = \tanh(W_h^ch_{t-1} + W_i^cx_t + b_c), \mbox{ } c_t = f_t\otimes c_{t-1} + i_t\otimes\widetilde{C}_t,
	\label{eq:candiate}
\end{equation}

where $b_c$ is a bias, and $\otimes$ indicates element-wise multiplication. The output gate $o_t$ regulates what to output to the next cell, \emph{i.e.,} the value of the next hidden state $h_t$ with operations shown in Eq.~\eqref{eq:output}.

\begin{equation}
	o_t = \sigma(W_h^oh_{t-1} + W_i^ox_t +b_o), \mbox{ } h_t = \tanh(c_t)\otimes o_t,
	\label{eq:output}
\end{equation}

where $b_o$ is a bias.

We formulate the task of adaptive EV position control for DWC as a multi-variate time series prediction problem, \emph{i.e.,} at any time step $t$, a future optimal position with maximum magnetic field strength is inferred based on the past optimal positions. More specifically, our objective is to predict future optimal points at time steps ($T_{t+1}$, $T_{t+2}, ..., T_{t+f}$), where $f$ is the future prediction horizon, based on past optimal points with highest electromagnetic strengths at time steps ($T_{t-l}$, $T_{t-l+1}$,...,$T_t$) where $l$ is the length of past observations. We set $f$ to 1 and $l$ to 10. We also note that these parameters can be easily changed depending on the requirements of applications.

\begin{figure}[h]
	\centering
	\includegraphics[width=.9\textwidth]{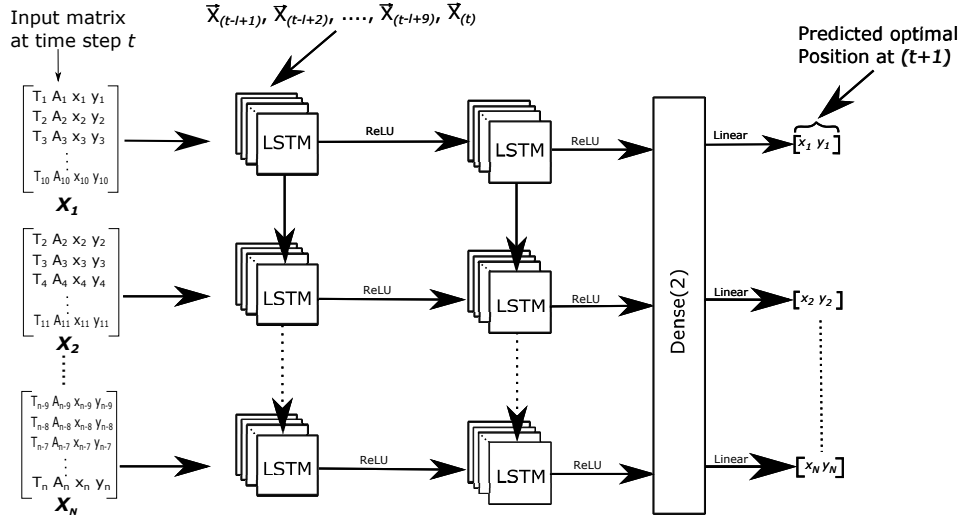}
	\caption {The architecture of the proposed multi-layer LSTM.}
	\label{fig:lstm_dwc}
\end{figure}

We design a multi-layer LSTM model based on the above-explained, basic LSTM network as a building block to solve this problem. By employing the multi-layer design, we aim to enable a more complex representation of time-series data and provide better prediction accuracy. Due to the computational overhead and limited hardware resources we have, we employ two layers in our design. 

Fig.~\ref{fig:lstm_dwc} displays the architecture of the proposed LSTM model. As shown, the input to train the LSTM model is a sequence of $4 \times 10$ matrices $\{X_1, ..., X_N\}$, each matrix containing 10 successive feature vectors. Essentially, an input matrix corresponds to a row of the feature matrix $D_{Features}$. More specifically, the 10 feature vectors of an input matrix, say at time step $t$, denoted by \{$\vec{x}^{(t-l+1)}$, $\vec{x}^{(t-l+2)}$, ...,$\vec{x}^{(t)}$\} are fed into the first LSTM layer. The first LSTM layer outputs a sequence of feature vectors as an input for the following LSTM layer, which is directly connected to the dense layer responsible for producing the $(p_x, p_y)$ coordinates of a point with maximum electromagnetic strength at time step $t+1$. Since predicting a 2D optimal position is a regression problem, the output of the fully connected layer is passed through the linear activation function. We train the model with mean squared error (MSE) loss function representing the difference between the ground truth point and predicted point: $MSE=\frac{1}{K}\sum_{i=1}^K(Y_i - \hat{Y_i})^2$, where $K$ is the number of the training sample, $Y_i$ is the actual value, and $\hat{Y_i}$ is the predicted value.

\begin{figure}[h]
	\centering
	\begin{minipage}[b]{0.47\textwidth}
		\centering
		\includegraphics[width=\linewidth]{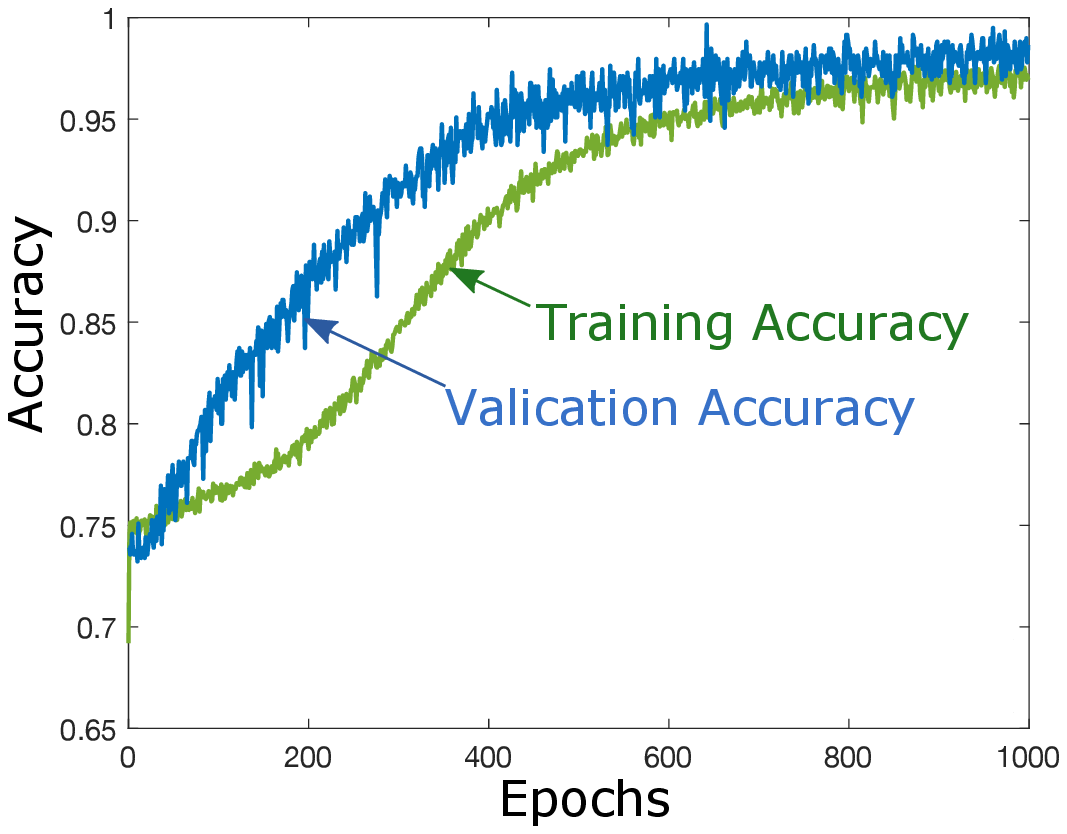}
		\caption {Training/Validation accuracy for the proposed LSTM model.}
		\label{fig:motivation}
	\end{minipage}
	\hspace{5mm}
	\begin{minipage}[b]{0.47\textwidth}
		\centering
	    \includegraphics[width=\linewidth]{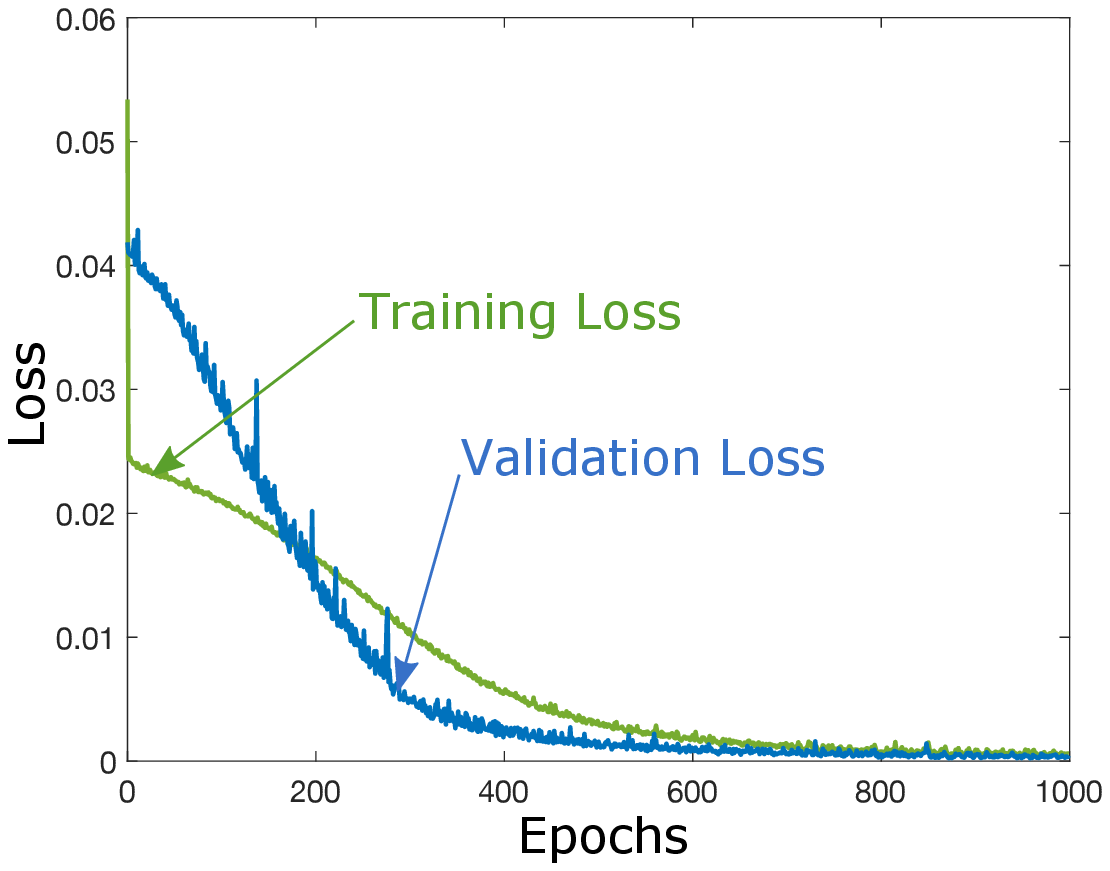}
	    \caption {Training/Validation loss for the proposed LSTM model.}
	    \label{fig:motivation2}
	\end{minipage}
\end{figure}

We perform training of the LSTM model for a sufficient number of epochs and use 10\% of our dataset for validating the training progress. Fig.~\ref{fig:motivation} and Fig.~\ref{fig:motivation2} show the accuracies and losses of our LSTM model, respectively.

%Although initially, the model exhibited poor performance on the validation dataset for hundreds of epochs, the performance gradually improves for both the training and validation set as training progresses. It achieves nearly the same performance for the training and validation dataset at the end of training. Therefore, the LSTM shows better prediction accuracy.

%The table \ref{tab:lstm_parameters} shows the network parameters used to train the LSTM model. We train the network with Adam optimizer with a learning rate of 0.0001 and a batch size of 64 to minimize the error between the target and predicted values. 

%\begin{table}[t]
%	\centering
%	\resizebox{0.9\textwidth}{!}{%
%		\begin{tabular}{|c|c|}
%			\hline
%			Number of Hidden Layers  &  2  \\ \hline
%			Hidden Units  &  100 in each Layer \\ \hline
%			Dense Layer  &  2 output Units \\ \hline
%			Activation  &  ReLu in hidden Layers, linear in output layer  \\ \hline
%			Loss Function & Mean Squared Error (MSE) \\ \hline
%			Optimizer & Adam (learning rate=0.0001)\\ \hline
%			Batch Size & 64 \\ \hline
%			
%		\end{tabular}%
%	}
%	\caption{LSTM network parameters}
%	\label{tab:lstm_parameters}
%\end{table}

%\subsection{Prediction}
%\label{sec:advanced_prediction}

%We assume that within t seconds the vehicle can position itself on the optimal location with the highest microelectric power. This t should be determined based on the width of a lane, type of the car, and speed of the car. For example, in this work, the literature shows that the max reposition rate is ... So we provide the output which represents the optimal location after t seconds in the future.

\section{Simulation Results}
\label{sec:results}

We evaluate our adaptive vehicle motion control system for DWC. We use the QuickField simulation software~\cite{quickfield} to simulate the transmitter and receiver coils of a regular DWC system. A point with maximum electromagnetic strength is measured with a time-step interval of 200ms. Specifically, a feature vector $\overrightarrow{x_i} = [T_i \mbox{ } A_i \mbox{ } p_x \mbox{ } p_y]$ is created at each time step $i$. A total of 6,000 feature vectors are created equivalent to about 20 minutes of DWC operation time to train the proposed LSTM model.

A PC equipped with M1 Processor with 8GB RAM running on Monterey OS is used for this simulation study. We implement, train, and test the proposed multi-layer LSTM model in Python using Keras and Tensorflow. An average training time for an epoch is 6 seconds, and we train our model up to 1000 epochs. Thus, the average training time for a model is approximately 1.6 hours. To measure the highest performance gain, we assume that the EV arrives at the LSTM-predicted optimal point in a timely manner. However, note that a more realistic scenario can be created by modeling vehicle arrival time as a stochastic process.

A key metric used for evaluation is the charging efficiency. With our test dataset, predictions are made a total of 600 times. The average electromagnetic strength of the predicted points is computed and used to represent the charging efficiency. The average electromagnetic strength of our approach is compared with state-of-the-art solutions focused on positioning the EV in the center of lane such as~\cite{tian2020vision}. We denote such a solution by the Base solution. In this section, we first determine the optimal hyperparameters for our multi-layer LSTM model (Subsection~\ref{sec:hyperparameter}). Based on the selected hyperparameters, we measure the charging efficiency by varying the dataset size and the prediction interval (Subsection~\ref{sec:dataset_size}). Finally, we measure the computational delay to ensure that the proposed approach is applicable for EVs with very high speed (Subsection~\ref{sec:computation_time}).

\subsection{Hyperparameter Tuning}
\label{sec:hyperparameter}

%\begin{figure}[h]
%	\centering
%	\includegraphics[width=.99\columnwidth]{tmp_table}
%	\caption {loss.}
%	\label{fig:motivation2}
%\end{figure}

We perform hyperparameter tuning to optimize the performance. The grid search method is used to find the optimal number of hidden units, batch size, and learning rate for our LSTM model. We consider the number of hidden units between 40 and 180 with an interval of 20, a set of batch sizes $\{16, 32, 64, 128, 256\}$, and a set of learning rates $\{0.01, 0.001, 0.0001, 0.00001, 0.000001\}$. The mean squared error (MSE) between the actual (ground truth) coordinates and the model-predicted coordinates is used to represent the effectiveness of hyperparameters. Consequently, for our simulation setting, we find the number of hidden units of 120, batch size of 16, and learning rate of 0.001 lead to optimal performance.

\subsection{Charging Efficiency}
\label{sec:dataset_size}

We analyze the charging efficiency of the proposed vehicle motion control system based on the set of hyperparameters that we find through hyperparameter tuning. Specifically, in our simulation, the charging efficiency is represented by the average electromagnetic strength computed over 600 times of predictions. We measure the average electromagnetic strength by varying the size of training dataset and prediction interval. Here the prediction interval determines how frequently the EV performs prediction and controls its motion accordingly to align itself with the optimal lateral position.

\begin{figure}[h]
	\centering
	\begin{minipage}[b]{0.47\textwidth}
		\centering
		\includegraphics[height=5.55cm, width=\linewidth]{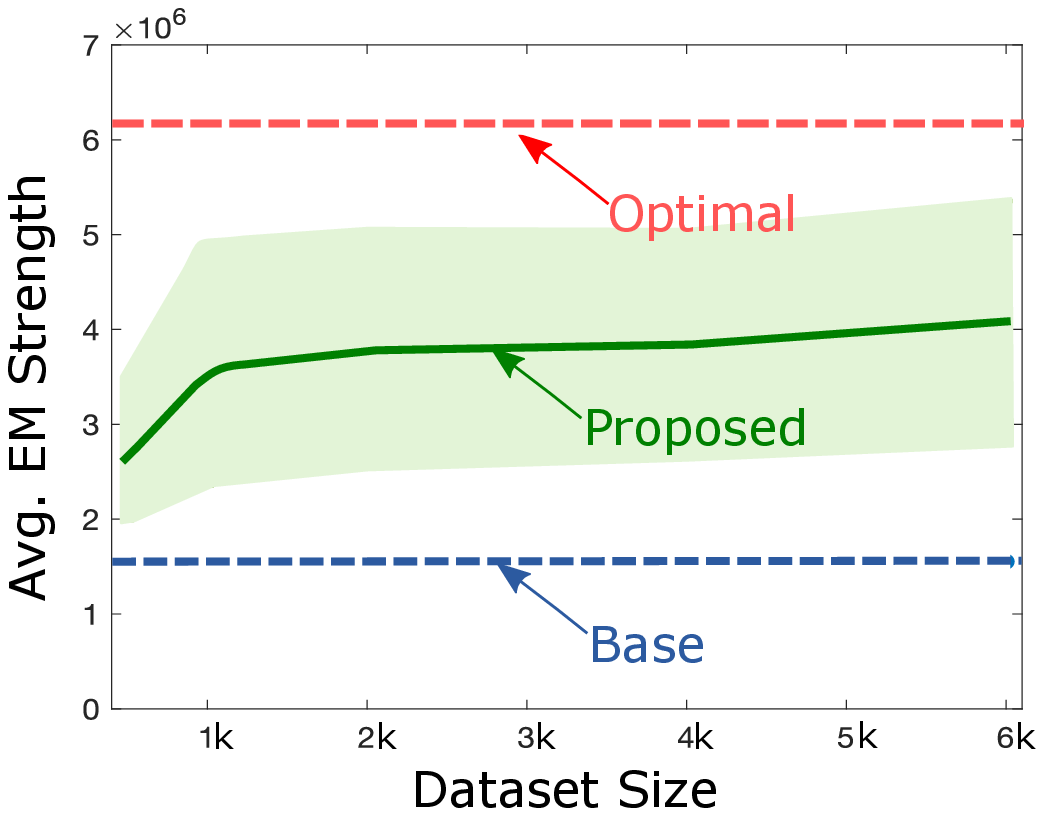}
		\caption {Charging efficiency with varying sizes of training dataset.}
		\label{fig:varyingdata}
	\end{minipage}
	\hspace{5mm}
	\begin{minipage}[b]{0.47\textwidth}
		\centering
		\includegraphics[width=\linewidth]{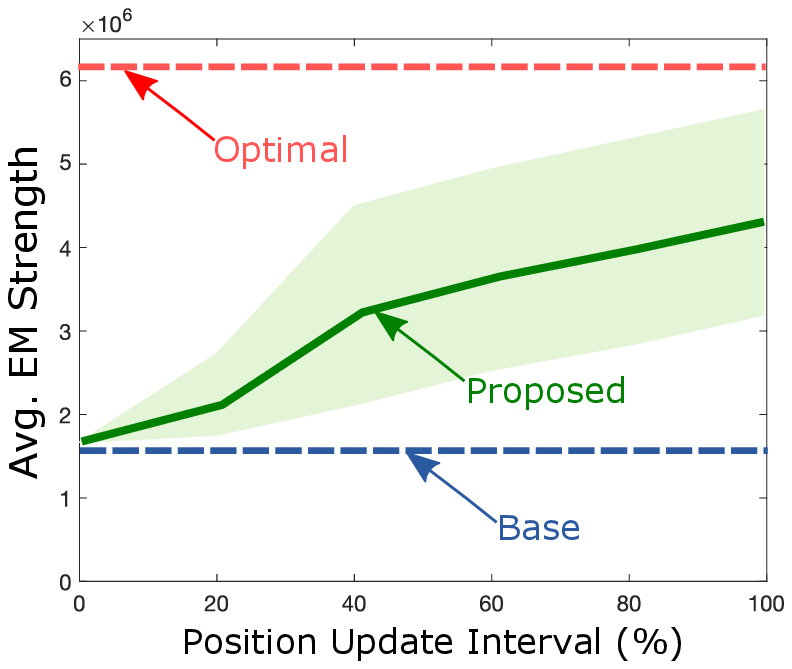}
		\caption {Charging efficiency with varying position update intervals.}
		\label{fig:updateinterval}
	\end{minipage}
\end{figure}

Fig.~\ref{fig:varyingdata} depicts the average electromagnetic strength with varying sizes of training dataset. It is observed that the average electromagnetic strength obtained with our solution is by up to 162.3\% higher than that of the Base approach. The results demonstrate the strong potential of dynamically adjusting the lateral position of EV to improve the charging efficiency even without modifying the existing infrastructure for DWC. Furthermore, as shown in Fig.~\ref{fig:varyingdata}, charging efficiency gradually improves as the size of training dataset increases. Specifically, we observe that with 6,000 records, the charging efficiency is improved by 48.2\% on average compared with that for 500 records. Fig.~\ref{fig:updateinterval} displays the average electromagnetic strength by varying the prediction interval. Here, 100\% interval means that the EV position is updated at its full capacity (determined based on the computational delay as explained in Section~\ref{sec:computation_time}).  The results demonstrate that higher charging efficiency can be achieved by making predictions and controlling the EV motion more frequently. It is notable that as the update interval is decreased to 20\%, the charging efficiency significantly decreases by 51.5\%. In particular, the proposed approach converges to the Base solution as the position update interval approaches 0\%. 

%\subsection{Update Interval}
%\label{sec:update_interval}

\subsection{Computational Delay}
\label{sec:computation_time}

\begin{wrapfigure}{r}{0.42\columnwidth}
	\vspace{-40pt}
	\begin{center}
		\includegraphics[width=\linewidth]{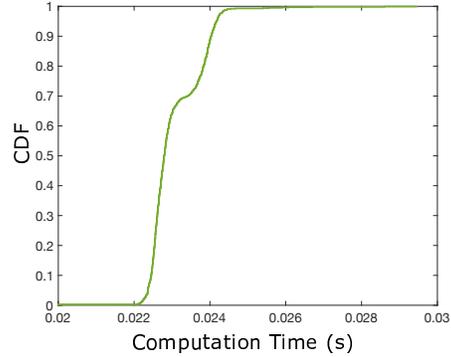}
		\caption{Cumulative distribution function (CDF) graph for computational delay.}
		\label{fig:computational_delay}
	\end{center}
\end{wrapfigure}

We have demonstrated in Section~\ref{sec:dataset_size} that more frequently updating the lateral position of EV leads to higher charging efficiency. One requirement for frequent position update is, however, sufficiently low computational delay for predicting the optimal lateral position. Fig.~\ref{fig:computational_delay} depicts the cumulative distribution function (CDF) graph for the computational delay for the proposed vehicle motion control system. The results demonstrate that in more than 95\% of time, the computational delay is smaller than 25ms, and the average computational delay is 23.8ms. The results indicate that with our system, a vehicle can adjust its lateral position every 8 to 10 meters on a highway when the vehicle speed is 75mph (120.7kmh).

\section{Conclusion}
\label{sec:conclusion}

We have presented a LSTM-based adaptive vehicle position control system for DWC. The dynamics of the electromagnetic field generated by the transmitter coils are effectively represented as a multi-layer LSTM model. The power transfer efficiency is improved by allowing an EV to predict the optimal point with maximum electromagnetic strength using the LSTM model and align its lateral position with the point. Simulation results demonstrate that our approach achieves by up to 162.3\% higher charging efficiency compared with a state-of-the-art solution focused on positioning the EV in the center of lane. 

Our future work is to address the limitation of the simulation-based evaluation by implementing a RC car-based prototype DWC system integrated with the proposed LSTM-based vehicle motion control solution to demonstrate significantly enhanced charging efficiency under realistic scenarios. Another future work is to address the limitation of sacalability by performing city-scale simulations to understand the large-scale, societal, and ecomic impact of the proposed vehicle motion control system for significant energy savings. A potential negative impact of fully automated EV motion control on driving comfort and safety for surrounding human-driven vehicles should also be addressed.

%%%%%%%%%%%%%%%%%%%%%%%%%%%%%%%%%%%%%%%%%%%%%%%%%%%%%%%%%%%%

\appendix

\bibliographystyle{aaai}
\bibliography{sample-base}

\end{document}